\documentclass[12pt,preprint]{aastex}
\begin{document}

\title{Long-term Optical Observations of  the Be/X-ray Binary X~Per}

\author{Hui Li\altaffilmark{1,2}, Jingzhi Yan\altaffilmark{1}, Jianeng Zhou\altaffilmark{1,2}, and Qingzhong Liu\altaffilmark{1}}

\altaffiltext{1}{Key Laboratory of Dark Matter and Space Astronomy, Purple Mountain Observatory, Chinese Academy of Sciences, Nanjing 210008, China}
\altaffiltext{2}{Graduate University of Chinese Academy of Science,Beijing 100049, China}
\email{jzyan@pmo.ac.cn}

\begin{abstract}
We present optical spectroscopic observations of X~Per from 1999 to 2013 with the 2.16m telescope at Xinglong Station and the 2.4m telescope at Lijiang Station, National Astronomical Observatories of China. Combining these observations with the public optical photometric data, we find certain epochs of anti-correlations between the optical brightness and the intensity of the H$\alpha$ and HeI 6678 lines, which may be attributed to the mass ejections from the Be star, however, alternative explanations are also possible. The variability of FeII~6317 line in the spectra of X~Per might also be caused by the shocked waves formed after the mass ejections from the Be star. The X-ray activities of the system might also be connected with the mass ejection events from the Be star. When the ejected materials were transported from the surface of the Be star to the orbit of the neutron star, an X-ray flare could be observed in its X-ray light curves. We use the neutron star as a probe to constrain the motion of the ejected material in the circumstellar disk. With the diffusion time of the ejected material from the surface of Be star to the orbit of neutron star, the viscosity parameter $\alpha$ of the circumstellar disk is estimated to be 0.39 and 0.28 for the different time, indicating that the disk around the Be star may be truncated by the neutron star at the 2:1 resonance radius and that a Type I X-ray outburst is unlikely to be observed in X~Per.

\end{abstract}

\keywords{stars: emission-line, Be---stars: individual (X~Per)---stars: neutron---X-rays: binaries}

\section{Introduction}
X Persei is a remarkable Be/X-ray binary that was first studied by the University of Michigan as part of its bright Be stars project that begin in 1913 \citep{Clark01}. X~Persei was later confirmed to be pulsar 4U 0352+30 \citep{braes72}. X Persei consists of a B0Ve star \citep{lyu97} and a compact companion that has a long pulse period $\sim$ 837 s \citep{hab98}, at a distance of 0.7--1.3 {\rm kpc} \citep{roc93,lyu97}. A 250.3 day orbital period and a low eccentricity of 0.11 were found from a pulse time analysis of the 837s pulsations by \citet{Del01}.

X Persei has highly variable optical and infrared emissions on timescales from minutes to years \citep{mook74,roc97}. The long-term photometry observations indicate that the optical brightness of X Per shows a large variation (V=6.1-6.8$^m$) \citep{roc93,Clark01}. The emission line profiles of H$\alpha$ and HeI 6678 {\AA} also display significant $V/R$ variability according to the long-term spectroscopic observations of \citet{Clark01}. A significant optical fading and the presence of absorbed photospheric H$\alpha$ and HeI 6678 {\AA} lines were observed between 1989 and1990 \citep{roc97}; these features indicated that X Per had lost its circumstellar disk and changed from a Be star to a normal B star. Recent optical spectroscopic observations by \citet{gru07} suggested that the size of the circumstellar disk around X Per was continuing to increaseand it showed a strong H$\alpha$ emission.

Due to its wide and low eccentric orbit, X Per is also distinguished by persistent, low-luminosity X-ray emission. An X-ray flare was detected in X Per between 1973 and1978 and a spin-up trend of $\dot{P}/P=-1.5\times10^{-4}~{\rm yr}^{-1}$ was also observed in the pulsation of the neutron star during that flare.After this, X~Per entered a spin-down stage in the last three decades, even through another small X-ray flare was observed around 2003 (see \citet{pal07} and references therein). A $\sim$29 keV cyclotron resonance scattering feature was found from the X-ray spectrum of X~Per by \citet{Coburn2001}. \citet{Lutovinov2012} reported the recent strong X-ray outburst activity of X~Per during 2001--2011 and identified a hard X-ray line around 160 keV in the X-ray spectrum for the first time.

In this work, we present the simultaneous visual photometric, spectroscopic, and X-ray observations in Section 2. We then analyze the variability of the H$\alpha$, HeI~6678, and FeII~6317 lines in our spectra, together with the long-term $V$-band light curves, and the activities of the X-ray emission of the system in Section 3. In Section~4, we discuss the relationship between the optical brightness and the intensity of the emission lines. We also study the correlation between the optical emission and the activities of the X-ray emission. At last, we give a summary of the main findings in Section~5.
\section{Observations}

\subsection{Optical Spectroscopy}
We have been annually monitoring the visual spectra of a number of Be/X-ray binaries since the 1990s. Most of X~Per's spectra were obtained with the 2.16 m telescope at Xinglong Station of National Astronomical Observatories of China. Optical spectroscopy with an intermediate resolution of 1.22\,{\AA} pixel$^{-1}$ was made with a CCD grating spectrograph at the Cassegrain focus of the telescope. We took the red spectra covering from 5500 to 6700\,{\AA} and blue spectra covering from 4300 to 5500\,{\AA} at different times. Sometimes low-resolution spectra (covering from 4300 to 6700\,{\AA}) were also obtained. In 2012 March and 2013 November, we also carried out spectroscopic observations with the Yunnan Faint Object Spectrograph and Camera (YFOSC) instrument of the Lijiang 2.4m telescope at Yunnan Astronomical Observatory. Grism \#8 was used with a resolution of 1.47\,{\AA} pixel$^{-1}$, with a spectral range from 5050 to 9750\,{\AA}. The data between 1992 and 1998 had been presented in \citet{liu01}. Here we only analyse the spectra between 1999 and 2013.
The journal of our observations is summarized in
Table~\ref{table:log}, including observational date, UT Middle, Julian Date, and spectral resolution. All spectroscopic data were reduced with the IRAF package\footnote{IRAF is distributed by NOAO, which is
operated by the Association of Universities for Research in Astronomy, Inc., under cooperation with
the National Science Foundation.}. They were bias-subtracted, flat-field corrected, and
had cosmic rays removed. Helium--argon spectra were taken in order to obtain the pixel-wavelength
relations.

The equivalent widths (EWs) of the H$\alpha$ line, the HeI 6678 line, and the FeII 6317, have been measured by selecting a continuum point on each side of the line and integrating the flux relative to the straight line between the two points using the procedures available in IRAF. The measurements were repeated five times for each spectrum and the error is estimated from the distribution of these values. The typical error for the measurements is within 1\%, arising due to the subjective selection of the continuum. The results of H$\alpha$, HeI 6678, and FeII 6317 EWs are listed in Table~\ref{table:log} and plotted in Figure~\ref{figure:xpermulti}. The published data of H$\alpha$ (adopted from \citet{Clark01}, \citet{liu01}, and \citet{gru07}) and HeI~6678 (adopted from \citet{Clark01}) are also plotted in Figure~\ref{figure:xpermulti} with open symbols.

\subsection{$V$-band Photometry}
We collect the $V$-band photometric data from \citet{roc93} and \citet{Clark01} and plot them in Figure~\ref{figure:xpermulti}. In order to study the optical brightness variability after 2001, we also make use of the public optical photometric data from the international database of the American Association of Variable Star Observers (AAVSO \footnote{http://www.aavso.org/}). Only the photometric observations made with the Johnson $V$-band filter are used  and plotted in Figure~\ref{figure:xpermulti}.

\subsection{X-Ray Light Curves}
The All Sky Monitor (ASM, 1.5--12~keV) on board \emph{RXTE} \citep{Levine96} had monitored the X-ray activity of X~Per from 1996 January to 2012 January. We plot the RXTE/ASM light curves rebinned with a time of 15 days in Figure~\ref{figure:xpermulti}. The Monitor of All-Sky X-Ray Image investigation \citep[MAXI;][]{Matsuoka2009} on the International Space Station (\emph{ISS}) also monitored the source since 2009 August in the 2--20~keV energy band. The MAXI light curves of X~Per are also plotted in Figure~\ref{figure:xpermulti} rebinned with a time of 15 days. We converted the observed count rates into the crab unit, by adopting 1 crab = 75 count~s$^{-1}$ for the ASM and 1 crab = 3.6 photons~s$^{-1}$~cm$^{-2}$  for MAXI, respectively.

\section{Results}
\label{Section:Results}

\subsection{Line Variability During our 1999--2013 Observations}
\subsubsection{H$\alpha$}
\begin{center}
\begin{figure}
\centering
\includegraphics[width=9cm]{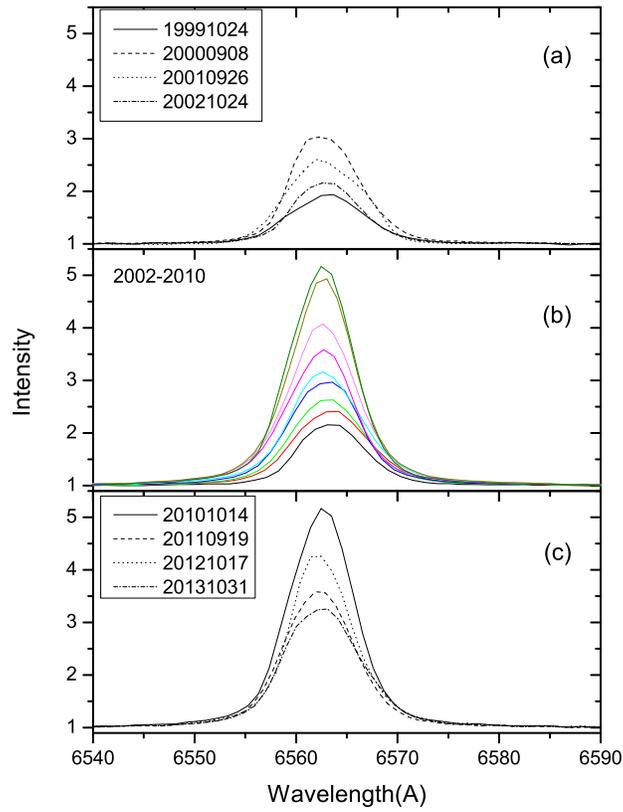}
\caption{H$\alpha$ profile evolution of X~Per from 1999 to 2013. (a) Lines between 1999 and 2002. (b) Lines between 2002 and 2010. The peak intensity kept increasing during our 2002--2010 observations. (c) Lines between 2010 and 2013. }
\label{figure:halpha}
\end{figure}
\end{center}

The evolution of the H$\alpha$ profiles during our 1999--2013 observations is shown in Figure~\ref{figure:halpha}. In our intermediate-resolution observations, all H$\alpha$ emission lines displayed a single-peaked structure from 1999 to 2013, while they showed a clearly double-peaked structure during our 1992--1995 observations \citep{liu01}.  The H$\alpha$ line profiles during 1999--2002 observations are plotted in Figure~\ref{figure:halpha}(a). The peak intensity of the H$\alpha$ emission line became stronger from 1999 October to 2000 September, and then kept declining in the following 2 yr, reaching another low value during our 2002 observations. Since 2002 October, the peak intensity of the H$\alpha$ line continued to increase until the observations in 2010 and these lines are displayed in Figure~\ref{figure:halpha}(b).   Even though the H$\alpha$ lines in our 1999 and 2002 observations have different peak intensities, they have nearly the same EW on average (see Table~\ref{table:log}). This might be caused by the broader wings in the 1999 observations than in the 2002 observations. With the increasing of the H$\alpha$ peak intensity after our 2002 observations, the wings of the H$\alpha$ also became increasingly broader and the H$\alpha$ line in 2010 is the broadest line during our 12 yr of observations. The H$\alpha$ lines during our 2010 and 2013 observations are shown in Figure~\ref{figure:halpha}(c), which indicates that the peak intensity of the line has an obvious decline in our 2011 observations. The peak intensity of the H$\alpha$ line became stronger again during our 2012 October observations and showed a lower level during our 2013 observations.

The H$\alpha$ EWs during our 1999--2013 observations  are plotted in Figure~\ref{figure:xpermulti}. During this period, the intensity of the H$\alpha$ showed a remarkable change. Its EWs varied between $\sim$ --6.5~\AA~and $\sim$ --37~\AA. The EWs of the H$\alpha$ showed a faint state in 1999 during our 15 yr observations. One year later, the H$\alpha$ EWs in 2000 were more than twice the 1999 averages. In the next 2 yr, the EWs of the H$\alpha$ line kept declining and reached another low level in our 2002 observations. After that, the H$\alpha$ EWs showed an increasing secular trend and a very strong H$\alpha$ emission line during our 2010 observations was observed with an EW of --37.06$\pm$0.10~\AA, which was the strongest in the past two decades. During our observations in 2011 September, the H$\alpha$ intensity showed an obvious decline and its EWs decreased to its 2007 level. Our observations with the 2.4m telescope at Lijiang Station in 2012 March indicated that the intensity of H$\alpha$ line was still in a decline stage. Interestingly, the H$\alpha$ intensity showed an increase during our observations in 2012 October.  In our 2013 observations, the intensity of the H$\alpha$ line decreased again and was at a level of that in 2012 March.

\subsubsection{HeI~6678}
\begin{center}
\begin{figure}
\centering
\includegraphics[width=9cm]{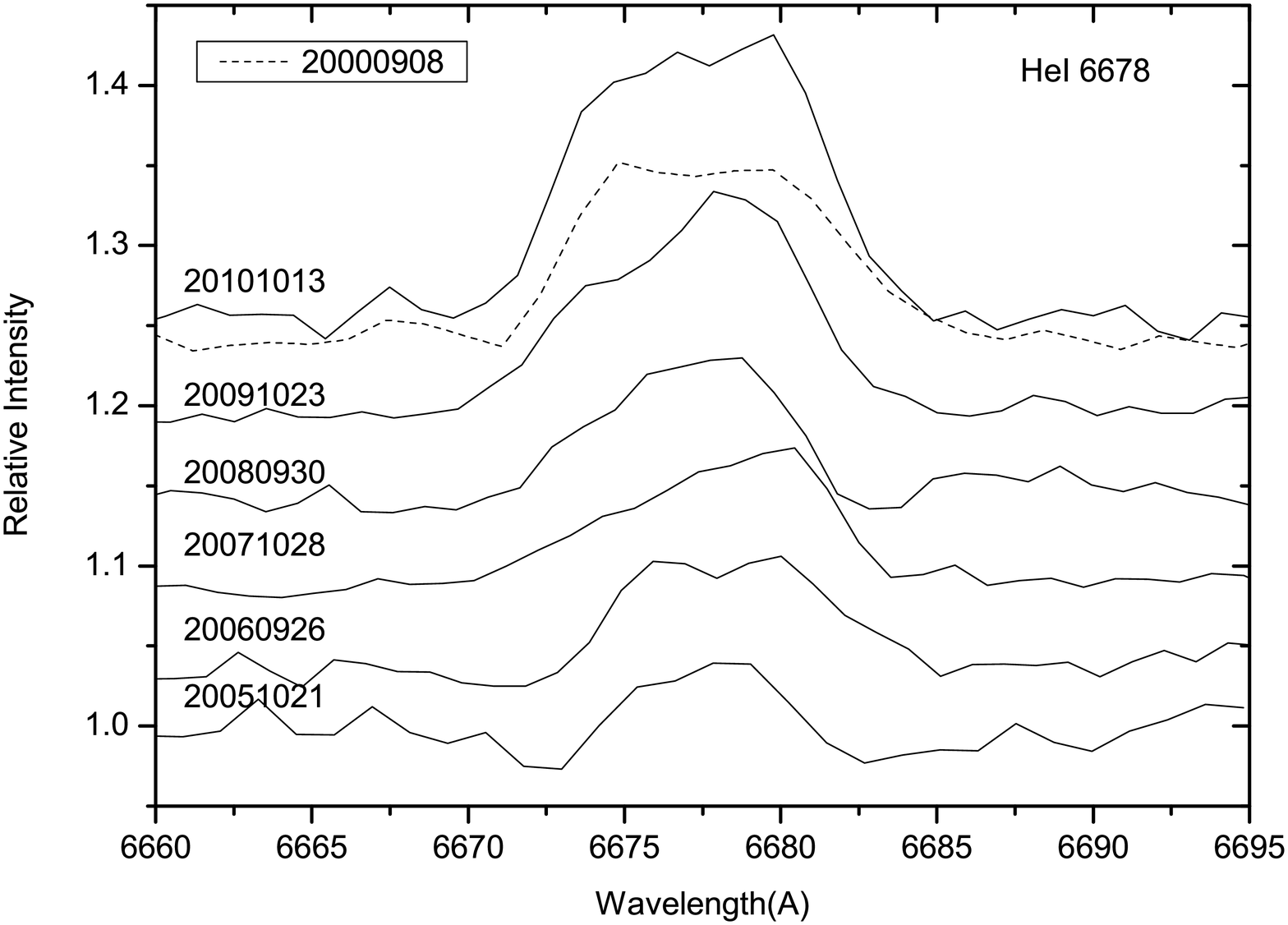}
\caption{HeI 6678 profiles during our 2005--2010 observations. The observational date is marked in the left part of each spectrum with a format of YYYYMMDD. The line observed on 2000 September 8 is also plotted in the figure with dashed lines.  }
\label{figure:hei}
\end{figure}
\end{center}

The HeI~6678 line was also observed in our spectra. Due to its faint emission state and/or bad pixels around 6678~\AA, we only plotted the HeI~6678 lines during the 2005--2010 observations in Figure~\ref{figure:hei}. Different from the H$\alpha$ lines plotted in Figure~\ref{figure:halpha}, the profile of HeI~6678 during these observations showed a double-peaked or asymmetrical single-peaked structure.  We also plot the HeI~6678 line during our 2000 observations in Figure~\ref{figure:hei} with a dashed line, , which indicates that the HeI~6678 line showed a strong emission and had an obvious double-peaked structure around 2000 September. The evolution of the HeI~6678 EWs is shown in Figure~\ref{figure:xpermulti}, which indicates that the HeI~6678 has a strong emission during the 2000 and 2001 observations and in the next 2 yr it nearly lost its emission in the spectra. After that the HeI~6678 emission showed a continuous increase over the following 3 yr. Unlike the H$\alpha$ line, the HeI~6678 showed a stable emission level during our 2006, 2007, and 2008 observations. It is interesting that the intensity of the HeI~6678 line kept increasing in the next two years and a strong HeI~6678 line was observed during our 2010 observations. It is worth noting that even though a relatively strong H$\alpha$ emission was observed during our 2010 observations, the HeI~6678 during 2010 observations had an intensity similar to that during 2000 observations. In the next few years, the HeI~6678 line has  the same evolutional mode as the H$\alpha$ line.

\subsubsection{FeII~6317}
\begin{center}
\begin{figure}
\centering
\includegraphics[width=9cm]{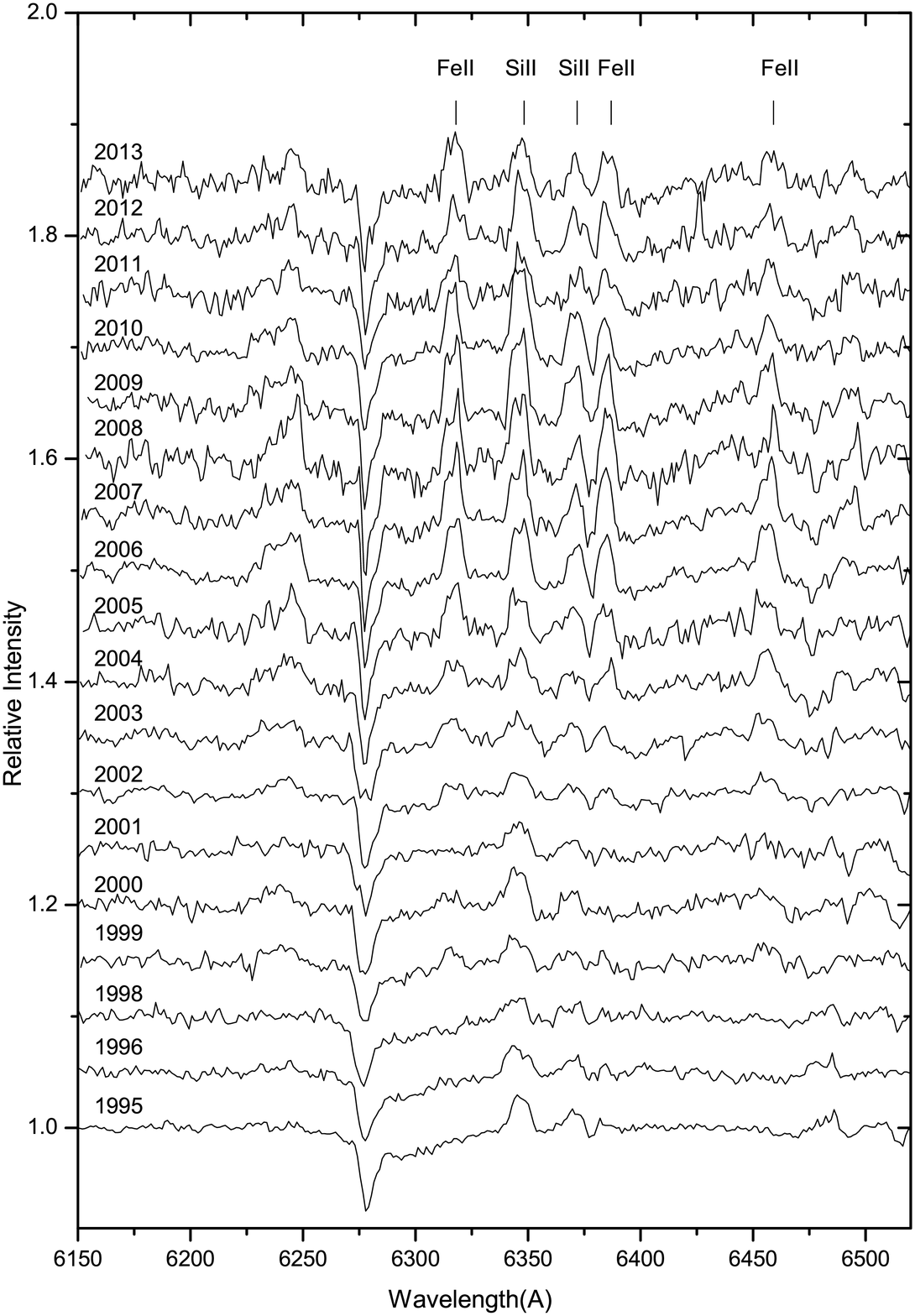}
\caption{Iron line profile evolution of X~Per from 1999 to 2013.}
\label{figure:fe}
\end{figure}
\end{center}

Several metal lines were also observed between 6200 {\AA} and 6500 {\AA} in the spectra.  We plot the spectra from 1995 to 2013 in Figure~\ref{figure:fe} and the metal lines are labeled in the top part of the figure. Among these lines, FeII~6317 is relatively simple. Here we take FeII~6317 as an example to study the evolution of the metal lines during our observations. Figure~\ref{figure:fe} indicates that the FeII~6317 emission feature  was nearly lost in our 1995, 1996, and 1998 observations. It appeared in our 1999 spectrum with an asymmetric profile and became faint in our 2000 observations with an obvious double-peaked structure. The emission of FeII~6317 was lost again during our 2001 observations. Since 2002, the FeII~6317 line has appeared again and its peak intensity has gradually increased sine then. The strongest FeII~6317 was observed during our 2009 observations. After 2009, the peak intensity of the FeII~6317 line began to decrease. Most of the FeII~6317 lines between our 2002--2013 observations showed an asymmetric double-peaked profile with a strong red peak except for the spectra during our 2012 observations, which showed a double-peaked structure with a strong blue peak. 

We also measure the EWs of the FeII~6317 line and plot them in Figure~\ref{figure:xpermulti}. Due to the faint emission of the FeII~6317 line and the low signal-to-noise ratio of the continuum around this region, the measure errors of the EW are large. Nevertheless, we can still see a continuous increase of the line intensity between 2002 and 2007. The intensity of the FeII~6317 line reamained stable in the following 2yr. Unlike the H$\alpha$ and HeI~6678 lines, the FeII~6317 intensity had an obvious decrease during our 2010 observations and was at a relatively low level during our 2011--2013 observations.

\subsection{Long-term Variabilities of the $V$-band Light Curves}
After a series of low-level fluctuations on timescales of $\sim$100 days, the $V$-band emission abruptly brightened by $\sim$ 0.6 mag within about 1 yr (MJD~51050--MJD~51383). The V-band brightness began to decrease when it reached a maximum ($V$$\sim$$6.2^m$) around MJD~51383. Between MJD~51383 and MJD~52165 (Region I in Figure~\ref{figure:xpermulti}(d)), the $V$-band brightness kept fading and reached another minimum ($V$ $\sim$ $6.6^m$) around MJD~52165. After that, the system showed another rapid increase in the $V$-band brightness, with a brighter state ($V$$\sim$ $6.1^m$) around MJD~52600. In the following 2 yr, the $V$-band emission kept a stable emission level with small fluctuations. 

Even for the sparse data between MJD~53383 and MJD~54107, we can still see a slow decrease of the $V$-band brightness. The $V$-band magnitude changed about $0.2^m$ within $\sim$2 yr. In the following year, the $V$-band brightness showed a slow increase and had a stable emission around MJD~54482.

Another obvious decrease in the $V$-band brightness was observed between MJD~54782 and MJD~55481 with an amplitude of $0.2^m$.  After MJD~55481, the $V$-band brightness had a rapid increase and a new fading phase of the optical brightness was observed around MJD~55990.

\subsection{Long-term X-Ray Activities}
 X-ray variability has been discussed by \citet{Lutovinov2012}. As shown in Figure~\ref{figure:xpermulti}, X~Per had stable X-ray emission at a low-level flux between MJD~$\sim$~50080 and $\sim$ 51999, with an average flux of $\sim$10 mcrab. The X-ray flux increased slowly after MJD~51999 and peaked around MJD~52781 ($\sim$ 50 m crab). After that, the flux declined rapidly to about 22~m crab within 150 days. The source stayed at another stable emission level with a higher X-ray flux for the following 5 yr and it began to brighten again around MJD~54640. When the source flux reached a new maximum around MJD~55363, it entered another quick decline phase. The new X-ray outburst  was also partly observed by MAXI in 2--20 keV. Due to the decommissioning of RXTE in 2012 January, small X-ray flares, peaking at MJD$\sim$55813 and MJD$\sim$56168, were only detected by MAXI.

\section{DISCUSSION}

\subsection{Correlations between the Visual Brightness and the Intensity of the Emission Lines}
\begin{center}
\begin{figure}
\centering
\includegraphics[width=10cm]{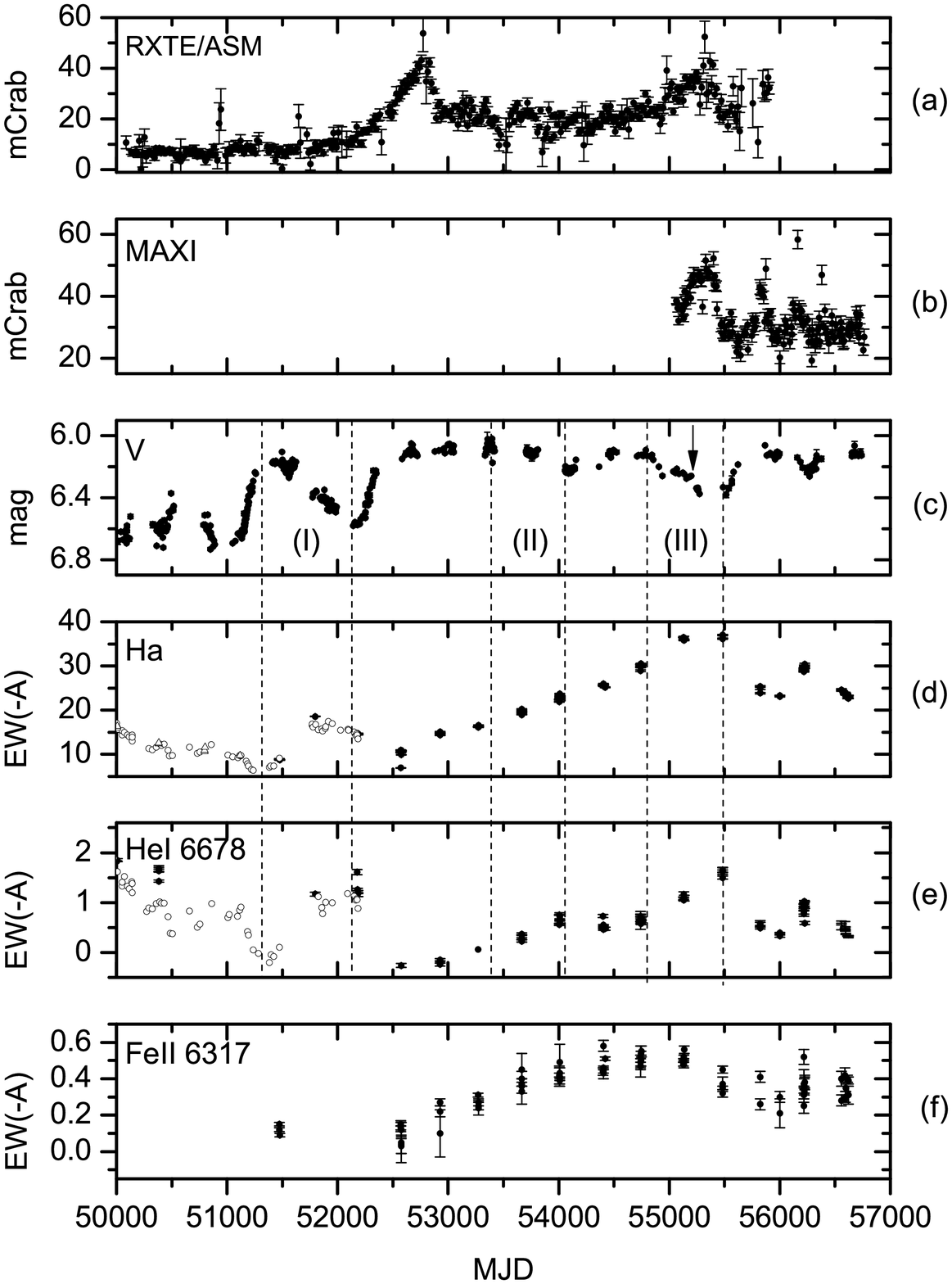}
\caption{Multi-wavelength observations of X~Per between MJD~50000 and MJD~57000. (a) The RXTE/ASM X-ray light curve in 1.5--12 keV rebinned with a time of 15 days. (b) The MAXI X-ray light curve in 2--20 keV rebinned with a time of 15 days. (c) The V-band photometric data adopted from  AAVSO. (d) The EWs of the H$\alpha$ line adopted from \citet{Clark01} (open circles), \citet{liu01} (open triangles),  and our observations during 1999--2013 (solid circles). (e) The EWs of the HeI~6678 line adopted from \citet{Clark01} (open circles) and our observations (solid circles). (f) The EWs of the FeII~6317 line of our observations. Three regions (I--III) are separated with dashed lines in (c), (d), and (e) panels.}
\label{figure:xpermulti}
\end{figure}
\end{center}

We have presented long-term optical photometric and spectroscopic observations of the Be/X-ray binary X Per in Section~\ref{Section:Results}. Several anti-correlations between the optical brightness and the intensities of the H$\alpha$ and the HeI~6678 lines are marked with three regions (I--III) separated by dashed lines in Figure~\ref{figure:xpermulti}.
When the visual brightness declined, the intensity of the H$\alpha$ and HeI~6678 lines showed an obvious increase. Before and after Region I, the opposite happened --when the visual brightness showed a rapid increase, the intensity of the emission lines decreased. A similar anti-correlation was also observed after Region III in Figure~\ref{figure:xpermulti}.

\begin{center}
\begin{figure}
\centering
\includegraphics[bb=65 35 726 508,height=10cm]{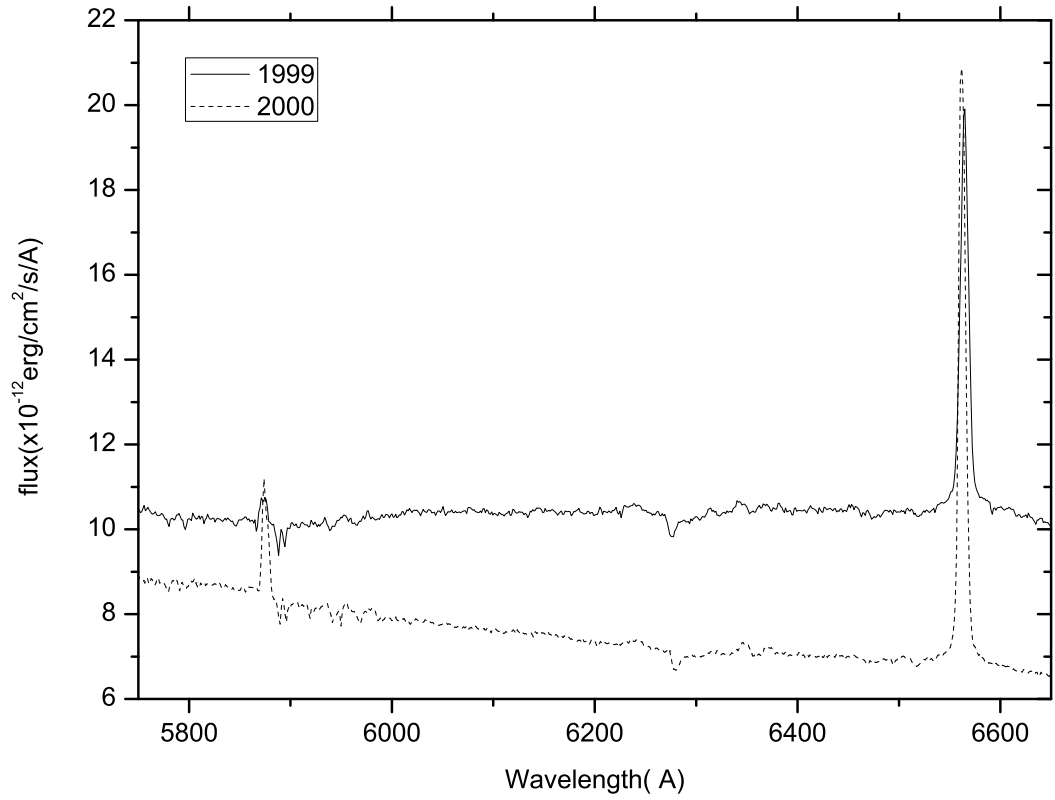}
\caption{Two typical continuum spectra selected from our 1999 and 2000 observations.}
\label{figure:continiuum}
\end{figure}
\end{center}

The increase in the line EWs might be caused by the decrease of the continuum emission. Two continuum spectra, selected from our 1999 and 2000 observations (the first two solid circles in Region I), are plotted in Figure~\ref{figure:continiuum}. Even through the spectrum in our 2000 observations had a lower continuum emission, it had stronger H$\alpha$ and HeI~5875 emissions than the spectrum in our 1999 observations. Therefore, we suggest that the increase in the line emission during Regions I--III in Figure~\ref{figure:xpermulti} might not be caused only by the decrease in the continuum emission; the physical changes in the circumstellar disc may also play an important role. To explain the anti-correlation between the optical brightness and the H$\alpha$ intensity, we should first know the physical origin of the optical continuum and H$\alpha$ emission in the system.

The optical thick H$\alpha$ emission line in a Be star is generally believed to be formed in the entire circumstellar disk \citep{Slettebak92}, while only the innermost part of the disk could affect the continuum flux \citep{Haubois2012}. Due to the higher ionization potential energy, the formation region of the HeI~$\lambda$6678 line should be smaller than the nearby continuum region \citep{Stee98}. The increase of the H$\alpha$ intensity indicates that a larger or denser circumstellar disk is forming around the Be star, while the decrease in the V-band brightness suggests that a low-density or cavitated region is developing in the inner part of the disk. Similar phenomena have also been observed in other Be/X-ray binaries, such as 4U~1145--619 \citep{Stevens97}, A0535+26 \citep{Clark99,Yan2012a}, 4U~0115+63 \citep{Reig07}, and MXB~0656--072 \citep{Yan2012b} . Observational results \citep{Rivinius01} and theoretical calculations \citep{Meilland06} suggest that after an outburst a low-density region seems to develop around the Be star. \citet{Rivinius01} and \citet{Meilland06} suggested that the outburst might be related to the increased mass loss or mass ejection from the Be star. Some weeks to months after the outburst, the stellar radiation pressure gradually excavates the inner part of the disk and a low-density region develops around the Be star and slowly grows outward \citep{Rivinius01}. Another possibility is that when the mass ejection from the Be star stops, a part of the ejected material would be reaccreted onto the surface of the star and a depleted region could be formed in the inner part of the disk \citep{Haubois2012}. With the vacation of the inner disk, the optical continuum emission decreases and an increase in $UBV$ magnitudes will be observed. After the outburst, material is transferred into the disk and a more extended circumstellar disk should form, which produces stronger H$\alpha$ emission from the system. The HeI~$\lambda$6678 became stronger when the optical continuum emission was decaying, indicating that a larger amount of material close to the stellar surface should be present when a low-density region is developing in the inner part of the disk. The new material around the surface of the star should be connected to the reaccreted material after the ejections \citep{Clark01}.

The decrease in the V-band brightness and the increase in the H$\alpha$ intensity during Regions I--III in Figure~\ref{figure:xpermulti} might be caused by the mass ejection from the Be star in X~Per. With the outward movement of the ejected materials, the size and the density of the disk kept increasing and a stronger H$\alpha$ emission was observed. A low-density region was formed in the inner region of the disk after the mass ejection, resulting in a continuous fading of the optical brightness. On the other hand, the outer part of the disk should also be truncated by the orbital motion of the neutron star in the system due to the gravitational interaction \citep{Okazaki01}. Therefore, in Figure~\ref{figure:xpermulti}, a stable H$\alpha$ emission was observed between MJD~$\sim$ 51772 and 52140 during Region I when the optical brightness was still in a fading phase. Since the HeI~6678 is formed in the surface of the Be star, the orbital motion of the neutron star has little effect on the inner part of the disk. This is the reason for a continuous increase of the HeI~6678 emission during Region I. The steady increase in the H$\alpha$ emission between our 2002 and 2010 observations might be, at least, connected with the mass ejection events that happened in Regions II and III.  With the decrease of the optical brightness during these two regions, the intensity of the H$\alpha$ and HeI~6678 lines showed an obvious increase. Between Regions II and III, the emission of the HeI~6678 line was stable, while the intensity of the H$\alpha$ line continued to increase. Once the mass ejection stopped after Region II, the material kept moving outward and the size of the circumstellar disk became larger and larger. This could explain the increase in the H$\alpha$ emission between these two regions. As a large mount of the materials were refilled after a series of mass ejections, a denser and more extended circumstellar disk formed around the Be star and a very strong H$\alpha$ emission was observed during our 2010 observations.

Region III in Figure~\ref{figure:xpermulti} likely consists of two different mass ejection events. The beginning time of the second fading is marked with an arrow lines around MJD~55215. A cavity region would be formed after the first mass ejection event and the second mass ejection would refill the depleted inner region of the circumstellar disk. With the outward movement of the cavity region, a double disk structure would be formed in the circumstellar disk of X~Per, which has been discussed by \citet{Tarasov95}, \citet{liu01}, and \citet{Clark01}. We do not rule out alternative possibilities for explaining the anti-correlations as have been proposed by \citet{Sigut2013} and \citet{Mathew2013}.

After Region~I and Region~III in Figure~\ref{figure:xpermulti}, the optical brightness of the system showed a rapid increase, which might mean that a new disk was being formed in the inner part of the circumstellar disk. Such a process may be weaker than the mass ejection events that occurred during Regions I--III, as we discussed above. Therefore, little material accumulated on the surface of the Be star and a rapid decrease in the HeI~6678 emission was observed after Regions I and III in Figure~\ref{figure:xpermulti}. The HeI~6678 emission feature was nearly lost during our 2002 observations. On the other hand, the circumstellar disk should be truncated by the orbital motion of the neutron star. Once the mass ejection stops, the size or the density of the circumstellar disk does not increase any longer and a rapid decrease in the H$\alpha$ emission is also observed after Regions I and III.

\subsection{The Physical Origin of the FeII Lines in X~Per}

FeII lines can be seen in the visual spectrum of many objects, while the FeII emission lines only appear in the classical Be stars earlier than B5 \citep{Hubert1979}. The ionization potential of neutral Fe is 7.8 eV, while that of FeII ions is 16.2 eV, which implies that FeII lines are formed in the regions close to the central star. Model calculations by \citet{Arias2006} indicate that the FeII lines that form in Be stars are optically thick and the extension of the line-forming region is 2.0 $\pm$ 0.8 times of the star radius.

All the FeII~6317 lines during our observations showed a double-peaked structure (see Figure~\ref{figure:fe}), while the H$\alpha$ line showed only a single strong peak during the same time, which also implies that the FeII lines in X~Per are formed in the inner part of the circumstellar disk. Figure~\ref{figure:xpermulti}(f) indicates that the FeII~6317 lines only appeared at the bright maximum. The intensity of the FeII~6317 had an continuous increase between our 2002 and 2006 observations and maintained a stable strong emission level between our 2006 and 2010 observations. During these time ranges, the intensity of the H$\alpha$ line also showed a continuous increase, which might be connected with serials of the mass ejection events from the Be star. Stronger shock waves could result after the mass ejections. The increase of the FeII~6317 emission might be also connected to the hot post-shock regions \citep{Richter2003}.

\subsection{The Connection between the X-Ray and Optical Activities}

The X-ray activities of the system should depend on the optical variability of the Be star. The neutron star in X~Per orbits the Be star with a low eccentricity ($e$ = 0.11) and a long orbital period ($P_{orb}$= 250.3 days). Therefore, X~Per shows a persistent X-ray source in the sky and its periodic X-ray outburst is not obvious in its X-ray light curve. No significance correlation has been found between the optical and X-ray light curves \citep{Lutovinov2012}.

The physical nature of the mass ejection from the Be star is currently still under debate. It is generally believed that the ejection might be connected with the rapid rotation of the Be star. The ejected material from the Be star seems to be very important for forming the circumstellar disk. The most widely accepted model for the circumstellar disk around a Be star is the viscous decretion disk model, first proposed by \citet{Lee1991}. In this model, the material is transported from the star outward according to the viscous diffusion timescale, $t_{\rm diffusion}~\propto~r^{1/2}/\alpha$, where $r$ is the distance to the central star, and $\alpha$ is the viscosity parameter first used by \citet{Shakura73} in the accretion disk model. It needs a long time, about several months to years, to transport the material from the surface of the Be star to the outer part of the circumstellar disk. Assuming that the neutron star X-ray emission is connected to the changes in its accreting environment, there should be a time lag between the activities of optical emission and the X-ray emission in a Be/X-ray binary system.

In a Be/X-ray binary system, the neutron star could be used as a probe for locating the diffused material in the circumstellar disk. As usual we assume that the mass ejection event took place at the beginning of the optical brightness fading, while the X-ray flux peaked when the densest part of the ejected material was transported to the orbit of the neutron star. We also assume that the peak time of the X-ray flare is the time of the periastron passage of the neutron star, since the orbital period is much less than the diffusion time of the ejected material in the circumstellar disk. The time for the beginning of the mass ejection event in Region I in Figure~\ref{figure:xpermulti} is $\sim$ MJD~51383, while the subsequent X-ray peak flux happened at $\sim$ MJD~52781. The time interval is about 1400 days.  Assuming an isothermal disk around the Be star, we can estimate the viscosity parameter $\alpha$  using Equation (19) of \citet{Bjorkman05},
\begin{equation}
\label{euqation1}
t_{\rm diffusion}=(0.2{\rm yr}/\alpha)*(r/R_*)^{0.5},
\end{equation}
where $t_{\rm diffusion}$, $\sim$ 1400~days, is the diffusion time of the ejected material from the surface of the Be star to the periastron point and $r$ is the radial distance from the Oe star to the periastron point of the neutron star. Given that the mass $M_*$ and the radius $R_*$ of the Be star X~Per are $15.5M_{\odot}$ and $6.5R_{\odot}$ \citep{gru07}, respectively, and the typical mass for a neutron star is $1.4M_{\odot}$, the value of $r/R_*$ is approximately 58.8, for an orbital period of 250.3~days with an eccentricity of $\sim$ 0.11. Therefore, the viscosity parameter $\alpha$ is $\sim$0.39 estimated by Equation~(\ref{euqation1}). Similarly, the second X-ray outburst that peaked around MJD~55363 might be caused by the mass ejection event that occurred around MJD~53383 in Region II in Figure~\ref{figure:xpermulti}. The diffusion time of $\sim$ 2000 days corresponds to a viscosity parameter of $\sim$~0.28 in Equation~(\ref{euqation1}).

The viscosity parameter $\alpha$ during the second X-ray outburst is smaller than the $\alpha$ during the first one. The smaller $\alpha$ suggests a lower radial diffusion velocity in the disk \citep{Okazaki2001,Lee1991} and it will take a longer time for the ejected material to flow from the surface of the Be star to the orbital of the neutron star. Moreover, the smaller $\alpha$ might be related to the higher density in the disk \citep{oka02} during the second X-ray outburst, when X~Per showed an extremely strong H$\alpha$ emission. These results suggests that the circumstellar disk around the Be star in X~Per may be truncated at the 2:1 resonance radius \citep{Clark01}, indicating that Type I X-ray outbursts are unlikely to be observed in X~Per.

\section{Conclusions}

We have presented our optical spectroscopic observations of X~Per from 1999 to 2013. Combining these with the public V-band brightness, we found several anti-correlations between the optical brightness and the strength of the H$\alpha$ and HeI~6678 lines: when the optical brightness was in a fading phase, the strength of the H$\alpha$ and HeI~6678 lines showed an obvious increase. This anti-correlation was interpreted as the result of the mass ejection event from the Be star. After mass ejection, a cavity or low-density region may develop in the inner part of the circumstellar disk. The H$\alpha$ emission exhibited a continuous increase during our 2002--2010 observations and it had the strongest intensity during our 2009 and 2010 observations.  The variability of the FeII~6317 line might be also caused by the shocked regions formed after the mass ejections events. The X-ray activities of X~Per might be caused by mass ejection events in the Be star. With the diffusion of the ejected material from the surface of the Be star to the orbit of the neutron star, X-ray outbursts could result in due to the increase in the mass accretion onto the neutron star.
The viscosity parameter $\alpha$ was estimated to be 0.39 and 0.28 from the different  diffuse times of the two mass ejection events taking place at MJD~51383 and MJD~ 53383, respectively. With these values, the circumstellar disk in X~Per may be truncated at the 2:1 resonance radius and the Type I X-ray outburst is unlikely to be observed in the system.

\acknowledgements We thank the anonymous referee and Anatoly Miroshnichenko for their valuable suggestions and Y. Z. Fan for his help with our manuscript. This work was partially supported by the Open Project Program of the Key Laboratory of Optical Astronomy, National Astronomical Observatories, Chinese Academy of Sciences; the 973 Program of China under grant 2013CB837000; the National Natural Science Foundation of China under grants 11003045, 11273064, 11443004; and the Strategic Priority Research Program "The Emergence of Cosmological Structures" of the Chinese Academy of Sciences, Grant No. XDB09000000. We also acknowledge the support of the staff of the Lijiang 2.4 m telescope. Funding for the  2.4 m telescope has been provided by CAS and the People's Government of Yunnan Province.

\begin{deluxetable}{lcccccc}
\tablecolumns{7} \tablecaption{Summary of the spectroscopic observations of X Per.}
\tablewidth{0pc}
\tablehead{\colhead{Date}&\colhead{UT} & \colhead{MJD} & \colhead{Spectral} &\colhead{H$\alpha$}  &\colhead{HeI6678} &\colhead{FeII 6317}\\
\colhead{(YYYYMMDD)}&\colhead{Middle} & \colhead{} & \colhead{Resolution} & \colhead{EW}  & \colhead{EW} & \colhead{EW}\\
\colhead{}&\colhead{(hh:mm:ss)} & \colhead{} & \colhead{(\AA~pixel$^{-1}$)} & \colhead{(-\AA)} & \colhead{(-\AA)}  & \colhead{(-\AA)}} \startdata
19991022 & 16:42:19 & 51473.696 & 1.22 & 8.85$\pm$0.06     & ...   & 0.15$\pm$0.01\\
19991022 & 16:44:49 & 51473.698 & 1.22 & 8.91$\pm$0.13   & ... & 0.12$\pm$0.01\\
19991024 & 16:01:49 & 51475.668 & 1.22 & 8.74$\pm$0.08 & ...   & 0.10$\pm$0.01 \\
20000908 & 19:43:27 & 51795.822 & 1.22 & 18.58$\pm$0.03 &1.17$\pm$0.04  & ... \\
20010925 & 18:22:14 & 52177.765 & 1.22 & 14.5$\pm$0.08 &1.61$\pm$0.05  & ... \\
20010926 & 18:15:04 & 52178.760 & 1.22 & 14.73$\pm$0.09 &1.26$\pm$0.03 & ... \\
20010930 & 18:37:03 & 52182.776 & 1.22 & 14.45$\pm$0.05 &1.18$\pm$0.06  & ... \\
20011001 & 19:11:05 & 52183.800 & 1.22 & 14.72$\pm$0.09 &1.19$\pm$0.01  & ... \\
20021023 & 18:03:31 & 52570.752 & 1.22 & 10.41$\pm$0.1 & ...   & ... \\
20021023 & 18:06:24 & 52570.754 & 1.22 & 6.93$\pm$0.06 & ...   & 0.14$\pm$0.03\\
20021023 & 18:09:11 & 52570.756 & 1.22 & 10.97$\pm$0.06 & ...  & 0.15$\pm$0.01 \\
20021024 & 19:06:14 & 52571.796 & 1.22 & 10.67$\pm$0.17 & ...  & 0.13$\pm$0.01\\
20021026 & 16:34:12 & 52573.690 & 1.22 & 9.92$\pm$0.04 & ...     & 0.12$\pm$0.04\\
20021026 & 16:38:07 & 52573.693 & 1.22 & 10.69$\pm$0.1 & ...      & ... \\
20021026 & 16:44:35 & 52573.698 & 1.22 & 10.6$\pm$0.08 & ...       & ... \\
20021027 & 19:27:40 & 52574.811 & 1.22 & 10.89$\pm$0.05 & ...      & ... \\
20021028 & 18:21:15 & 52575.765 & 1.22 & 10.89$\pm$0.05 &-0.26$\pm$0.04  & ... \\
20031014 & 18:25:44 & 52926.768 & 1.22 & 14.89$\pm$0.1 &-0.24$\pm$0.03   & 0.22$\pm$0.03  \\
20031014 & 18:27:37 & 52926.769 & 1.22 & 15.04$\pm$0.15 &-0.19$\pm$0.02  & ...\\
20031015 & 19:53:45 & 52927.829 & 1.22 & 14.74$\pm$0.05 &-0.15$\pm$0.03   & 0.27$\pm$0.02\\
20031016 & 17:17:23 & 52928.720 & 1.22 & 14.41$\pm$0.13 & ...    & ... \\
20031016 & 17:18:07 & 52928.721 & 1.22 & 14.35$\pm$0.06 & ...     & ... \\
20040921 & 19:04:42 & 53269.795 & 1.22 & 16.19$\pm$0.06  &0.06$\pm$0.01  & 0.31$\pm$0.01\\
20040925 & 19:32:08 & 53273.814 & 1.22 & 16.55$\pm$0.1  & ...    & 0.25$\pm$0.02\\
20040926 & 18:51:54 & 53274.786 & 1.22 & 16.27$\pm$0.02  & ...  & 0.24$\pm$0.04\\
20040926 & 18:50:34 & 53274.785 & 1.22 & 16.21$\pm$0.15  & ...   & 0.27$\pm$0.02\\
20051021 & 20:05:11 & 53664.837 & 1.22 & 19.46$\pm$0.07  &0.23$\pm$0.02   & 0.4$\pm$0.04\\
20051023 & 17:29:27 & 53666.729 & 1.22 & 19.21$\pm$0.11 & ...     & 0.37$\pm$0.01\\
20051023 & 20:04:31 & 53666.836 & 1.22 & 19.92$\pm$0.14 &0.28$\pm$0.02   & 0.36$\pm$0.04\\
20051023 & 20:05:30 & 53666.837 & 1.22 & 18.97$\pm$0.06 &0.23$\pm$0.02    & ... \\
20051024 & 18:41:22 & 53667.779 & 1.22 & 19.81$\pm$0.05  &0.35$\pm$0.01  & 0.33$\pm$0.07\\
20051024 & 18:42:27 & 53667.779 & 1.22 & 20.23$\pm$0.12  &0.37$\pm$0.02  & 0.45$\pm$0.09\\
20060926 & 19:57:10 & 54004.831 & 1.22 & 22.4$\pm$0.11    &0.56$\pm$0.02   & 0.42$\pm$0.05\\
20060927 & 18:44:17 & 54005.781 & 1.22 & 22.42$\pm$0.05  &0.57$\pm$0.02  & 0.42$\pm$0.01 \\
20060928 & 18:19:49 & 54006.764 & 1.22 & 21.9$\pm$0.1 &0.75$\pm$0.01       & 0.41$\pm$0.05\\
20060929 & 18:16:01 & 54007.761 & 1.22 & 22.61$\pm$0.12 &0.66$\pm$0.01    & 0.43$\pm$0.03\\
20061001 & 18:25:33 & 54009.768 & 1.22 & 23.09$\pm$0.17 &0.61$\pm$0.02    & 0.49$\pm$0.10\\
20061001 & 18:27:07 & 54009.769 & 1.22 & 23.06$\pm$0.09  &0.65$\pm$0.02    & 0.41$\pm$0.05\\
20061002 & 19:00:51 & 54010.792 & 1.22 & 23.68$\pm$0.08   &0.72$\pm$0.08   & 0.39$\pm$0.02\\
20071028 & 20:28:13 & 54401.853 & 1.22 & 25.67$\pm$0.09   &0.73$\pm$0.03   & 0.44$\pm$0.02\\
20071030 & 20:46:37 & 54403.866 & 1.22 & 25.71$\pm$0.05   &0.50$\pm$0.02   & 0.43$\pm$0.03\\
20071031 & 20:49:43 & 54404.868 & 1.22 & 25.48$\pm$0.12 &0.46$\pm$0.01    & 0.58$\pm$0.03\\
20071101 & 19:15:20 & 54405.802 & 1.22 & 25.75$\pm$0.11  &0.51$\pm$0.01   & 0.44$\pm$0.01\\
20071101 & 19:16:59 & 54405.803 & 1.22 & 25.95$\pm$0.13 &0.55$\pm$0.01    & 0.46$\pm$0.01\\
20071116 & 16:34:37 & 54420.691 & 1.22 & 25.21$\pm$0.09 &0.48$\pm$0.03     & 0.51$\pm$0.01\\
20080930 & 19:30:44 & 54739.813 & 1.22 & 30.41$\pm$0.13 &0.73$\pm$0.02     & 0.49$\pm$0.02\\
20081001 & 17:51:36 & 54740.744 & 1.22 & 29.77$\pm$0.18 &0.70$\pm$0.07     & 0.47$\pm$0.02\\
20081001 & 20:32:31 & 54740.856 & 1.22 & 30.02$\pm$0.16 &0.64$\pm$0.18     & 0.52$\pm$0.01\\
20081002 & 20:32:31 & 54741.856 & 1.22 & 28.91$\pm$0.18  &0.65$\pm$0.07    & 0.47$\pm$0.06\\
20081002 & 18:42:23 & 54741.779 & 1.22 & 29.98$\pm$0.38  &0.64$\pm$0.03    & 0.52$\pm$0.06\\
20081005 & 18:08:13 & 54744.756 & 1.22 & 30.5$\pm$0.11 &0.65$\pm$0.02       & 0.54$\pm$0.01\\     
20081006 & 18:07:18 & 54745.755 & 1.22 & 29.88$\pm$0.2 &0.58$\pm$0.02       & 0.50$\pm$0.03\\
20081009 & 19:04:54 & 54748.795 & 1.22 & 30.17$\pm$0.09 &0.68$\pm$0.03     & 0.55$\pm$0.03\\
20091023 & 20:21:46 & 55127.848 & 1.22 & 36.49$\pm$0.08  &1.08$\pm$0.03    & 0.48$\pm$0.01\\
20091026 & 19:05:49 & 55130.796 & 1.22 & 35.84$\pm$0.14  &1.16$\pm$0.03    & 0.48$\pm$0.02\\
20091028 & 19:46:38 & 55132.824 & 1.22 & 36.44$\pm$0.28  &1.05$\pm$0.02    & 0.52$\pm$0.01\\
20091029 & 18:54:44 & 55133.788 & 1.22 & 35.89$\pm$0.11  &1.09$\pm$0.06     & 0.56$\pm$0.02\\
20101012 & 20:00:59 & 55481.834 & 1.22 & 37.06$\pm$0.10   &1.65$\pm$0.05    & 0.32$\pm$0.02\\
20101013 & 20:16:09 & 55482.844 & 1.22 & 36.16$\pm$0.07   &1.50$\pm$0.03    & 0.45$\pm$0.02\\
20101014 & 16:45:14 & 55483.698 & 1.22 & 36.31$\pm$0.05   &1.63$\pm$0.02    & 0.37$\pm$0.04\\
20101015 & 18:54:50 & 55484.788 & 1.22 & 36.98$\pm$0.11   &1.64$\pm$0.07     & 0.35$\pm$0.01\\
20110917	&18:39:01	 & 55821.777	&	1.22	&	23.87$\pm$0.14	&	0.49$\pm$0.02 	    & ... \\
20110918	&18:36:43	 & 55822.776	&	1.22	&	25.37$\pm$0.16	&	0.57$\pm$0.03 	    & 0.41$\pm$0.03\\
20110919	&18:13:40	 & 55823.760	&	1.22	&	25.02$\pm$0.33	&	0.58$\pm$0.06 	    & 0.26$\pm$0.03\\
20120313$^{*}$ & 13:15:36 &	55999.552	&	1.47	&	23.1$\pm$0.14	&	0.39$\pm$0.01     & 0.21$\pm$0.08 	\\
20120315$^{*}$ & 12:15:15 &	56001.511	&	1.47	&	23.3$\pm$0.05	&	0.33$\pm$0.02 	            & 0.30$\pm$0.03\\
20121016	& 19:59:31 &	56216.833	&	1.22	&	28.92$\pm$0.25	&	0.95$\pm$0.03   & 0.37$\pm$0.08	\\
20121017	& 17:37:40 &	56217.734	&	1.22	&	28.92$\pm$0.1	&	0.86$\pm$0.02         & 0.31$\pm$0.04 	\\
20121017	& 21:04:01 &	56217.878	&	1.22	&	28.75$\pm$0.08	&	0.94$\pm$0.02  & 0.25$\pm$0.04 	\\
20121018	& 17:49:12 &	56218.742	&	1.22	&	29.17$\pm$0.13	&	0.78$\pm$0.05  & 0.34$\pm$0.02 	\\
20121019	& 18:17:34 &	56219.762   &	1.22	&	29.53$\pm$0.09	&	0.81$\pm$0.04 	         & 0.52$\pm$0.04\\
20121021	& 20:18:48 &	56221.846	&	1.22	&	29.68$\pm$0.19	&	0.91$\pm$0.02  & 0.36$\pm$0.04	\\
20121022	& 20:14:03 &	56222.843	&	1.22	&	29.96$\pm$0.26	&	1.02$\pm$0.02 	 & 0.33$\pm$0.02\\
20121024	& 18:52:50 &	56224.787	&	1.22	&	29.62$\pm$0.15	&	0.59$\pm$0.03  & 0.38$\pm$0.03	\\
20121025	& 17:41:34 &	56225.737	&	1.22	&	30.40$\pm$0.17	&	0.95$\pm$0.07  & 0.42$\pm$0.04 	\\
20130921 &17:22:50  &  56556.724    &	1.22& 24.57 $\pm$ 0.09 & 0.57 $\pm$ 0.02 & 0.28 $\pm$ 0.03 \\
20130923 &18:05:30  & 56558.754    &	1.22& 24.57 $\pm$ 0.32 & 0.54 $\pm$ 0.09 & 0.4 $\pm$ 0.04 \\
20131025 &17:20:01  & 56590.722   &	1.22& 23.97 $\pm$ 0.13 & 0.55 $\pm$ 0.07 & 0.42 $\pm$ 0.04 \\
20131026 & 18:33:10  & 56591.773     &	1.22& 23.95 $\pm$ 0.07 & 0.49 $\pm$ 0.02   & ... \\
20131027 &17:03:36   & 56592.711    &	1.22& 23.88 $\pm$  0.09 & 0.44 $\pm$  0.02 & 0.30 $\pm$  0.03 \\
20131029 & 18:31:06   &56594.772      &	1.22& 23.5 $\pm$ 0.06 & 0.47 $\pm$ 0.04 & 0.29 $\pm$ 0.02 \\
20131031 & 16:10:29   &56596.674  &	1.22 & 23.59 $\pm$ 0.21 & 0.37 $\pm$ 0.02 & 0.35 $\pm$ 0.02 \\
20131118$^{*}$ &14:55:43  & 56614.622       &	1.47	& 22.78 $\pm$ 0.2 & 0.33 $\pm$ 0.01 & 0.4 $\pm$ 0.02 \\
20131123$^{*}$ & 15:04:47 &56619.628      &	1.47	& 23.32 $\pm$ 0.11 & 0.31 $\pm$ 0.01& 0.31 $\pm$ 0.05 \\
20131124$^{*}$  & 14:24:16 &56620.600   &	1.47	& 22.69 $\pm$ 0.09 & 0.34 $\pm$ 0.01 & 0.39 $\pm$ 0.02 \\

\enddata
\label{table:log}
\tablenotetext{Note:}{The data marked with an asterisk after the observational date are obtained with the Lijiang 2.4 m telescope.}

\end{deluxetable}


\begin{thebibliography}{}


\bibitem[Arias et al.(2006)]{Arias2006}
Arias et al., 2006, A\&A, 460, 821

\bibitem[Bjorkman \& Carciofi(2005)]{Bjorkman05}
Bjorkman, J.E., \& Carciofi, A.C., 2005, in ASP Conf. Ser. 337, The Nature and Evolution of Disks Around Hot Stars, ed. R. Ignace \& K. Gayley (San Francisco, CA: ASP), 75

\bibitem[Braes et al.(1972)]{braes72}
Braes, L. L. E. \& Milley, G.K.  1972, Nature, 235, 273


\bibitem[Clark et al.(1999)]{Clark99}
Clark, J.S., Lyuty, V.M., Zaitseva, G.V., et al., 1999, MNRAS, 302, 167

\bibitem[Clark et al.(2001)]{Clark01}
Clark, J. S., Tarasov, A. E., Okazaki, A. T., Roche, P., \& Lyuty, V.M.   2001, A\&A, 380, 615

\bibitem[Coburn et al.(2001)]{Coburn2001}
Coburn W., Heindl W. A., Gruber D. E., Rothschild R. E., Staubert R., Wilms J., Kreykenbohm I., 2001, ApJ, 552, 738

\bibitem[Delgado-Mati et al.(2001)]{Del01} Delgado-Marti, H., Levine, A. M., Pfahl, E., \& Rappaport, S. A. 2001, ApJ, 546, 455

\bibitem[Grundstrom et al.(2007)]{gru07} Grundwtrom, E. D., Boyajian, T. S., Finch, C. et al. 2007, ApJ, 660, 1398
\bibitem[Haberl et al.(1998)]{hab98} Haberl, F., Angelini, L., Motch, C., White, N. E. 1998, A\&A, 330, 189


\bibitem[Haubois et al.(2012)]{Haubois2012}
Haubois, X., Carciofi, A.C., Rivinius, Th., Okazaki, A.T., and Bjorkman, J. E., 2012, \apj, 756, 156


\bibitem[Hubert-Delplace \& Hubert(1979)]{Hubert1979}
Hubert-Delplace, A. M., \& Hubert, H. 1979, An Atlas of Be Stars, Paris Meudon Observatory
\bibitem[Lee et al.(1991)]{Lee1991}
Lee, U., Saio, H., Osaki, Y., 1999, MNRAS, 250, 432

\bibitem[Levine et al.(1996)]{Levine96}
Levine, A. M., Bradt, H., Cui, W., Jernigan, J. G., et al., 1996, ApJ, 469, L33

\bibitem[Liu \& Hang(2001)]{liu01} Liu, Q. Z., \& Hang, H. R. 2001, Ap\&SS, 275, 401
\bibitem[Lutovinov et al.(2012)]{Lutovinov2012}
Lutovinov, A., Tsygankov, S., and Chernyakova, M., 2012, MNRAS, 423, 1978

\bibitem[Lyubimkov et al.(1997)]{lyu97} Lyubimkov, L. S., Rostopchin, S. I., Roche, P. \& Tarasov, A. E. 1997, MNRAS, 286, 549

\bibitem[Matsuoka et al.(2009)]{Matsuoka2009}
Matsuoka, M. et al., 2009, PASJ, 61, 999


\bibitem[Mathew et al.(2013)]{Mathew2013}
Mathew, Blesson, Banerjee, D. P. K., Naik, Sachindra and Ashok, N. M., 2013, AJ, 145, 158

\bibitem[Meilland et al.(2006)]{Meilland06}
Meilland, A., Stee, Ph., Zorec, J., and Kanaan, S., 2006, \aap, 455, 953
\bibitem[Mook et al.(1974)]{mook74} Mook, D. E., Boley, F. I., Foltz, C. B. et al. 1974, PASP, 86, 894

\bibitem[Okazaki(2001)]{Okazaki2001}
Okazaki, A. T., 2001, PASJ, 53, 119

\bibitem[Okazaki et al.(2002)]{oka02} Okazaki, A. T., Bate, M. R., Ogilvie, G. I., \& Pringle, J. E.  2002, MNRAS, 337, 967

\bibitem[Okazaki \& Negueruela(2001)]{Okazaki01}
Okazaki, A. T. and Negueruela, I., 2001, \aap, 377, 161



\bibitem[Palombara \& Mereghetti(2007)]{pal07} Palombara, N. L., Mereghetti, S. 2007, A\&A, 474, 137

\bibitem[Reig et al.(2007)]{Reig07}
Reig, P., Larionov, V., Negueruela, I., et al., 2007, \aap, 462, 1081

\bibitem[Richter et al.(2003)]{Richter2003}
 Richter et al., 2003, A\&A, 400, 319

\bibitem[Rivinius et al.(2001)]{Rivinius01}
Rivinius, Th., Baade, D., Stefl, S., and Maintz, M., 2001, \aap, 379, 257

\bibitem[Roche et al.(1993)]{roc93} Roche, P., Coe, M. J., Fabregat, J., et al. 1993, A\&A. 270, 122
\bibitem[Roche et al.(1997)]{roc97} Roche, P., Larionov, V., Tarasov, A. E., et al. 1997, A\&A, 322, 139

\bibitem[Shakura \& Syunyaev(1973)]{Shakura73}
Shakura, N.I., \& Syunyaev, R.A., 1973, \aap, 24, 337

\bibitem[Sigut \& Patel(2013)]{Sigut2013}
Sigut, T. A. A. \& Patel, P., 2013, \apj, 765, 41

\bibitem[Slettebak et al.(1992)]{Slettebak92} Slettebak, A., Collins, G.W., II, Truax, R., 1992, ApJS, 81, 335
\bibitem[Stee et al.(1998)]{Stee98} Stee, Ph., Vakili, D., Bonneau, D., \& Mourard, D., 1998, \aap, 332, 268
\bibitem[Stevens et al.(1997)]{Stevens97} Stevens, J.B., Reig, P., Coe, M.J., et al., 1997, MNRAS, 288, 988

\bibitem[Tarasov \& Roche(1995)]{Tarasov95}
Tarasov, A. E., \& Roche, P. 1995, MNRAS, 276, L19

\bibitem[Yan et al.(2012a)]{Yan2012a}
Yan, J.Z., Li, H., and Liu, Q.Z., 2012a, ApJ, 744, 37
\bibitem[Yan et al.(2012b)]{Yan2012b}
Yan, J.Z., Chaty, S., Zurita Heras, J.A., Li, H., \& Liu, Q.Z., 2012b, ApJ, 753, 73

\end{thebibliography}
\end{document}